# Real-time terahertz near-field microscope

F. Blanchard[1,2,*], A. Doi[2,3], T. Tanaka[1,2], H. Hirori[1,2], H. Tanaka[2,4], Y. Kadoya[2,4], and K. Tanaka[1,2,5]

[1] *Institute for Integrated Cell-Material Sciences, Kyoto University, Sakyo-ku, Kyoto 606-8501, Japan*

[2] *CREST, Japan Science and Technology Agency, Kawaguchi, Saitama 332-0012, Japan*

[3] *Olympus Corporation 2-3 Kuboyama-cho, Hachioji-shi, Tokyo 192-8512, Japan*

[4] *Department of Quantum Matter, ADSM, Hiroshima University, 1-3-1 Kagamiyama, Higashi-Hiroshima 739-8530, Japan*

[5] *Department of Physics, Graduate School of Science, Kyoto University, Sakyo-ku,* Kyoto 606-8502, Japan

blanchard@icems.kyoto-u.ac.jp

**Terahertz (THz) waves have been significantly developed in the last fifteen years because of their great potential for applications in industrial and scientific communities[1,2]. The unique properties of THz waves as transparency for numerous materials and strong absorption for water-based materials are expected to broadly impact biosensing[3] such as medical imaging[4], chemical identifications[5], and DNA recognition[6]. In particular, for accessing information within a scale comparable to the cell size where interactions between cell membrane and other organelle structures occur, micron size spatial resolution is required. Due to the large wavelength, however, the joint capability of THz near-field imaging with real-time acquisition, which is a highly desirable ability for observing real-time changes of *in vivo* sample, remains undone. Here, we report a real-time THz near-field microscope with a high dynamic range that can capture images of a 370 x 740 $\mu m^2$ area at 35 frames per second. We achieve high spatial resolution on a large area by combining two novel techniques: THz pulse generation by tilted-pulse-front excitation[7] and electro-optic (EO) balanced imaging detection using a thin crystal. To demonstrate the microscope capability, we reveal the field enhancement at the gap position of a dipole antenna after the irradiation of a THz pulse. Our results are the first demonstration of a direct quantification of a 2-dimensional subwavelength THz electric field taken in real-time.**

The very first demonstration of THz imaging[8] clarified that microscopic resolution is required for applications ranging from semiconductors inspections[9] to biomedical imaging[3,4]. For a corresponding wavelength at 1 THz, however, the far-field spatial resolution of an image is restricted to the Rayleigh criterion that corresponds to ~180 μm in vacuum. To overcome this limit, we must work in the near-field region of a sample. In this case, the THz wave has to be captured very close to the sample surface in a region before light diffraction occurs[10]. Many schemes have been proposed since the first demonstration of THz near-field imaging[11], most of which involve a subwavelength probe or a aperture placed very near the sample[9,11-17].

However, even if spatial resolution below 10 μm is successfully achieved at THz frequency, all of the traditional schemes remain based on raster scanning techniques[9,11-17]. In such cases, the imaging acquisition time associated with optical scanning techniques is relatively long and quadratically increases as a function of area size (i.e., a few hours for a 600 x 600 μm scan with 20-μm spatial resolution[12]). Ideally, to map in real-time the morphological changes of such subwavelength structures as cells or microorganisms, a new approach with a faster acquisition time has to be developed.

In this work, we demonstrate how to capture large THz near-field images with high dynamic range and a high-speed acquisition rate through the recent progress in the generation of intense THz waves[7]. In particular, images of a 370 x 740 μm² area are captured at a 35-Hz repetition rate with a spatial resolution of 14 μm (corresponding to λ/30 based on a center frequency of 0.7 THz). In the supplementary information no.1, we provide a movie of the reemitted THz electric field evolution of a metallic dipole antenna after irradiation by an intense and uniform THz pulse. The spectroscopic information, available for each pixel of the camera, successfully quantifies the THz electric field enhancement inside the 3-μm gap position of the dipole antenna and agrees well with simulations.

The THz microscope, which is divided into three parts (THz source, sample observation, and analyzer part), is coupled to a Ti: Sapphire laser beam line that delivers 4 mJ pulses with a 85-fs pulse duration at a 1-kHz repetition rate. In method section, experimental setup shown in Fig. 1 is described.

**Figure 1**

One key point for preserving the high spatial resolution of the optical elements when dealing with THz images is the merit of using a very thin EO crystal to collect the THz field from a sample before diffraction. This operation can be understood if we remember that the overlapping between the probe light and the THz waves in the EO crystal forms a THz image. On the point of view of the probing light, the EO crystal thickness must be selected to compare with the depth of the field range of the objective lens: ~20 μm in our case. For current THz emitter and detector technologies based on ultra-fast mode-locked lasers, however, which are commonly used for raster scanning THz microscopes, it is impossible to generate the high intensity THz images required to compensate for the lost encounter with thin EO sensors and preserve a high dynamic range over a large area. The decreased sensitivity caused by the choice of a very thin EO sensor[18] must be compensated by an intense THz wave source. On the other hand, one advantage of using a thin EO sensor is the possibility of working with LN material that is transparent for visible light and exhibits a larger EO coefficient when compared to the traditionally used ZnTe material[19].

Figure 2(a) shows the visible and THz images of a metallic mask directly deposited on top of the EO sensor with their corresponding extracted profiles in Fig. 2(b). We evaluated the spatial resolution of the microscope by taking the time-resolved spatial distribution of the electric field of an edge varying from 10% to 90% in amplitude[20]. We found a spatial resolution of 14 μm with a 20-μm-thick LN x-cut crystal. The extracted visible profile of Fig. 2(b) confirms that the optical elements do not restrict the THz spatial resolution. Detailed analysis of the spatial resolution will be reported in a different paper.

**Figure 2**

At the development state of our microscope and before imaging the biological samples, we must improve our knowledge of the imaging of subwavelength metallic structures. As previously reported, numerous studies from visible to THz frequencies ranges have investigated various subwavelength structures[12,21-27], such as single hole and array, to understand more about strongly localized fields. In particular, resonator structures, which are good candidates to quantify electric field distribution, are expected to facilitate the enhancement of the field at the point of interaction with the sample[22,28,29], thus improving the sensitivity. To demonstrate our microscope's performance, we measured the THz electric field in the near-field region of a metallic dipole antenna after the irradiation of a uniform THz pulse. In Fig. 3(a) and (b), we show a schematic description and a visible image of the sample, respectively. The dipole antenna composed of two adjacent (73.5 × 10.0 × 0.2 μm) gold bars and forming a 3-μm air gap was directly fabricated on the LN crystal. The antenna was oriented parallel to the polarization of the incoming pulse ($E_{THz}$) and set to maximize the EO modulation. In this particular geometry, angle $\theta$ of 38° is formed between the *Y*-axis of the LN crystal and a direction normal to the antenna (Fig. 3(b) and (c)).

**Figure 3**

Figure 3(d) shows six temporal snapshots of the electric field passing through and reemitted by a dipole antenna. Since the LN material exhibits a large birefringence ($k_y \neq k_z$), the reemitted THz waves from the dipole antenna will experience a shift ($\phi = \sim10°$) following the $\boldsymbol{k}_{THz}$ direction that can be described by the phonon-polariton[29] properties of the traveling THz waves inside the LN crystal (Fig. 3(c)). To visualize the temporal evolution of the THz electric field inside and around the dipole antenna after the irradiation of an intense THz pulse, we presented 175 movie frames of the field as supplementary information. This movie was captured in about nine minutes of acquisition and covers 9.2 ps of temporal evolution with 53 fs in step resolution. All images were obtained from 100 averaged images taken at a 35-Hz repetition rate. Figures 3(e) and (f) are the Fourier transformations obtained at each pixel position with their time-dependent field distribution yields spectrally resolved in amplitude (e) and phase (f), respectively. The movies of the amplitude and phase information for the different frequency components are also shown as supplementary material no.2 and no.3, respectively. Note the interesting spatial beating frequency that appears in the Fourier domain of Figs. 3(e) and (f). This phenomenon can be explained by the Fourier transformation of the main pulse with the pulse reemission from the dipole antenna. This situation can be described by the following expressions:

$$E(x,\omega) = \int_{-\infty}^{\infty} E(x,t) \cdot e^{-i\omega t} dt \qquad (1),$$

$$E(x,t) = E_{THz}(t) + E_{Ant}(t - \frac{x}{v(\omega)}) \qquad (2),$$

where $x$ is the distance from the dipole antenna and $v(\omega)$ is the speed of the THz waves inside the LN material as a function of THz frequency $\omega$. For a one-dimensional case, function $E(x,t)$ represents the total electric field including the THz pump excitation $E_{THz}$ and the reemitted field from the antenna $E_{Ant}$. In the presence of two pulses whose phase difference changes with location $x$ inside the LN crystal, a rectified periodic signal is observed. In fact, the highest amplitude of the beating frequency appears around the enhancement frequency of the antenna, where both amplitudes ($E_{THz}$ and $E_{Ant}$) are almost equivalent in our case.

To confirm that our microscope can quantitatively measure subwavelength electric fields, we compared the simulated enhancement factor with the experimental measurements. We extracted the time-dependent field distribution for each pixel below the gap, which corresponds to an area of 3 pixels x 7 pixels, and performed the Fourier transformation of this corresponding area. In particular, since the SNR of any given pixel (where one pixel corresponds to an area of 1.4 x 1.4 $\mu m^2$) is 100, we could perform the spectroscopy at the gap position of the dipole antenna with an SNR greater than 400. Note again this microscope's ability to perform local field measurements at any pixel position from the CCD camera.

**Figure 4**

In Fig. 4(a), we show the Fourier spectrums of the THz pulses with and without an antenna for the measured and simulated THz pulses. The simulated enhancement spectrum obtained in Fig. 4(a) is an average value of the enhancement spectrums shown in Fig. 5(c) (as described in the method section). Fig. 4(b) compares the experimental and simulation data obtained from the ratios of the enhancement field with the reference spectrums. There is a small difference between the simulation and experimental results. It may come from an error between the fabricated antenna and the one used in the simulations. In addition, the simulations were made on an isotropic and thick medium, neglecting the LN crystal birefringence that can lead to an underestimation of the real effective refractive index of the LN crystal. Nevertheless, this demonstration agrees well with the simulations and shows that we can quantitatively measure the THz electric field in a near-field of a sample and in real-time.

**Figure 5**

In conclusion, we introduced a new type of THz near-field microscope that combines a high intensity THz pulse source with a thin EO crystal sensor. This system can retrieve large images in real-time with high spatial resolution and a high dynamic range. In particular, we achieved a spatial resolution of 14 μm ($\lambda$/30 at 0.7 THz) with a 370 x 740 μm$^2$ image size taken at a 35-Hz repetition rate. Our results also demonstrate the ability to quantitatively map the electric field enhancement at the gap position of a dipole antenna along with good agreement with FDTD simulations.

Further improvements in spatial resolution are expected using a thinner EO crystal. Moreover, by selecting a larger imaging sensor the imaging area can be increased without compromising the spatial resolution. Finally, we stress the real-time capability of our microscope, which in a few minutes can take the equivalent of 22.9 million single data point acquisitions to form the full length THz near-field movie shown as supplementary information. We believe this microscope has immense potential for applications that require real-time tracking changes of *in vivo* microstructures such as cells or microorganisms.

**Method**

**Experimental setup**

The generation part is based on a novel technique of optical rectification in LiNbO$_3$ by tilted-pulse-front excitation[7]. By assuming a Gaussian profile, we evaluated the THz beam spot size diameter and the peak electric field at the focus to be 0.5 mm and 200 kV/cm. For detection of the pulse THz waveform, we performed free-space EO sampling[18] in an x-cut LiNbO$_3$ (LN) crystal with a thickness of 20 μm, mounted on a 0.5-mm-thick glass, which could be reconstructed by using an optical delay stage. To observe the near-field region, the sample is placed directly on top of the EO crystal[19]. Notice that the top and bottom surfaces of the EO crystal have high-reflection and anti-reflection coating for the probe light at 800 nm. As shown in Fig. 1, the probe light at 800 nm forms an image of the iris on the top surface of the EO crystal after passing through a 200-mm focal length achromatic lens in combination with an 18-mm focal length objective lens. The reflected images are returned to the objective lens and retrieved by a CCD camera with a non-polarized beamsplitter cube. To transfer the image to the camera, a second 200-mm focal length achromatic lens is used; note that the camera is placed 200 mm from this lens to fulfill the imaging condition. The measurement of the THz-induced birefringence in the EO crystal is detected by a combination of a quarter-wave plate, a half-wave plate, and two polarizing beamsplitter cubes, as depicted in the analyzer part of Fig. 1. In addition, a balanced imaging scheme was used to improve the signal to noise ratio (SNR) and the imaging acquisition speed. This operation was done by spatially separating and simultaneously capturing the vertically and horizontally polarized probe images by one camera (Hamamatsu C9100-12 with 512 x 512 pixels). All corresponding pixels (from both the S and P polarized images) are subtracted in real-time, giving 35 background-free images per second, with a size of 256 x 512 pixels and depth of 14 bits.

## Simulation parameters

To quantitatively validate the electric field of the microscope and to get further insight into the physics involved, simulations were carried out using finite difference time domain (FDTD) software. The simulations were performed on a 101 x 215 μm area with a spatial resolution of 1 x 1 μm. The dipole antenna formed by two adjacent (73.5 × 10.0 × 0.2 μm) gold bars with a 3- μm air gap was oriented following the Y-axis of an x-cut 1-mm-thick LN crystal whose. The index of the LN crystal was set to 5.11[30], ignorneglecting the birefringence of the LN material. . The dielectric parameters of the gold metal[26] were used to perform the Drude model:.

$$\varepsilon(\omega) = \varepsilon_\infty - \frac{\omega_p^2}{\omega(\omega + i\gamma)} \qquad (3)$$

where $\varepsilon_\infty = 1$, the plasma frequency is $\omega_p = 1.37$ x 10$^4$ THz, and $\gamma = 40.7$ for gold. The exciting THz pulse has a center frequency at 1 THz and a full width half maximum (FWHM) of 1.0 ps.

In Fig. 5(a), we show the electric field mapping of the dipole antenna obtained 1 μm below the LN crystal surface. In the simulation, we can clearly see the electric field distribution inside the 3-μm-width gap with the 1 x 1 μm mesh resolution.

**Evaluation of simulated enhancement factor**

As mentioned previously, the microscope's imaging resolution is $\lambda/30$ and is strongly affected by the EO crystal thickness. To understand more about how the imaging of the field is affected as a function of the *z* direction, we simulated the fields for three different depths: 1, 10, and 20 μm. In addition, to adequately simulate the microscope behavior, we applied a Gaussian low pass filter with a 7-μm radius to lower the resolution of the simulated fields to ~14 μm spatial resolution. Fig. 5(b) shows the electric field mapping for the three layers below the LN crystal surface.

To estimate enhancement factor $\alpha(\omega)$, we used the ratio in amplitude with and without the antenna at the position of the antenna gap. Fig. 5(c) shows the Fourier transformed of the synthesized incoming reference THz pulse with the Fourier transformed of the field below the gap for the three corresponding depths. By looking at the spectrums shown in Fig. 5(c), the field enhancement can be found at a frequency of ~0.5 THz. To understand the distribution of the field enhancement inside the LN crystal, we present in Fig. 5(d) the enhancement field at 0.5 THz as a function of the *z* direction. The blue and red lines represent the simulations obtained with and without spatial filtering. As a first approximation, the electric field enhancement measured by the microscope can be estimated by the average value of the different components of the blue curve obtained in Fig. 5(d). Therefore, the dotted lines shown in Fig. 5(d) are the average values of the enhancement field factor integrated completely through the crystal depth. We estimate the enhancement factor to be ~2.6 at a frequency of 0.5 THz (represented by the blue dotted line), which is in good agreement with the experimental enhancement factor found in Fig. 4(b).

**Acknowledgements:** This study was supported by a Grant-in-Aid for Scientific Research and from MEXT of Japan (Grants Nos. 18GS0208 and 20104007). F. B. wish to thanks the FQRNT (Le Fonds Québécois de la Recherche sur la Nature et les Technologies (FQRNT) contract no. 138131.

**Author contributions:** K. T. and H. H. initiated and provided management oversight for this project. F. B. and A. D. designed and characterized the experiments. F. B. also carried out the measurements and analyzed the data. T. T. and A. D performed the simulations. H. T. and Y. K. designed and fabricated the metallic structures. All authors contributed to the final manuscript.

**Additional information**

The authors declare no competing financial interests.

**References**

[1] Ferguson, B. & Zhang, X.-C. Materials for terahertz science and technology. Nature Mater. **1**, 26–33 (2002).

[2] Horiuchi, N. Terahertz technology: Endless applications. Nature Photon. **4**, 140 (2010).

[3] Siegel, P. H. Terahertz technology in biology and medicine. Microwave Symposium Digest. IEEE MTT-S International, **3**, 1575-1578 (2004).

[4] Fitzgerald, A. J., et al. An introduction to medical imaging with coherent terahertz frequency radiation. Phys. Med. Biol. 47 R67-R84 (2002).

[5] Mittleman, D. *Sensing with Terahertz Radiation* (Springer, 2003).

[6] Markez, A. G., Roitberg, A. & Heilweil, E. J. Pulsed terahertz spectroscopy of DNA, bovine sérum albumin (BSA) and collagen between 0.1 and 2.0 terahertz. Chem. Phys. Lett., **320**, 42-48, (2000).

[7] Fülöp, J. A, Pálfalvi, L., Almási, G., & Hebling, J. Design of high-energy terahertz sources based on optical rectification. Opt. Express **18**, 12311-12327 (2010).

[8] Hu, B. B. & Nuss, M. C. Imaging with terahertz waves. Opt. Letters, **20**, 1716-1718, (1995).

[9] Huber, A. J., Keilmann, F., Wittborn, J., Aizpurua, J. & Hillenbrand, R. Terahertz Near-Field Nanoscopy of Mobile Carriers in Single Semiconductor Nanodevices. Nano. Lett. **8**, 3766 (2008).

[10] Dereux, A., Girard, C. & Weeber, J.-C., Theoretical principles of near-field optical microscopies and spectroscopies, J. Chem. Phys. **112**, 7775-7789 (2000).

[11] Hunsche, S., Koch, M., Brener, I. & Nuss, M. C. Terahertz near-field imaging. Opt. Commun. **150**, 22 (1998).

[12] Bitzer, A. & Walther, M. Terahertz near-field imaging of metallic subwavelength holes and hole arrays. Appl. Phys. Lett. **92**, 231101-231103 (2008).

[13] Chen, Q., Jiang, Z., Xu, G. X. & Zhang, X.-C. Near-field terahertz imaging with a dynamic aperture technique. Opt. Lett. **25**, 1122 (2000).

[14] van der Valk, N. C. J. & Planken, P. C. M. Electro-optic detection of subwavelength terahertz spot sizes in the near field of a metal tip. Appl. Phys. Lett. **81**, 1558-1560 (2002).

[15] Chen, H.-T., Kersting, R. & Cho, G. C. Terahertz imaging with nanometer resolution. Appl. Phys. Lett. **83**, 3009-3011 (2003).

[16] von Ribbeck, H.-G., et al. Spectroscopic THz near-field microscope. Opt. Express **16**, 3430-3437 (2008).

[17] Knab, J. R., Adam, A. J. L., Chakkittakandy, R. & Planken, P. C. M. Terahertz near-field microspectroscopy. Appl. Phys. Lett. **97**, 031115 (2010).

[18] Wu, Q. & Zhang, X.-C. Free-space electro-optic sampling of terahertz beams. Appl. Phys. Lett. **67**, 3523-3525 (1995).

[19] Winnewisser, C., Uhd Jepsen, P., Schall, M., Schyja, V. & Helm, H. Electro-optic detection of THz radiation in LiTaO$_3$, LiNbO$_3$ and ZnTe, Appl. Phys. Lett. **70**, 3069-3071 (1997).

[20] Doi, A., Blanchard, F., Hirori, H. & Tanaka, K. Near-field THz imaging of free induction decay from a tyrosine crystal. Opt. Express **18**, 18419-18424 (2010).

[21] Liedberg, B., Nylander, C. & Lundström, I. Biosensing with surface plasmon resonance – how it all started. Biosensors Bioelectron. **10**, i-ix (1995).

[22] Ebbesen, T. W., Lezec, H. J., Ghaemi, H. F., Thio, T. & Wolff, P. A. Extraordinary optical transmission through sub-wavelength hole arrays. Nature **391**, 667-669 (1998).

[23] Lezec, H. J., et al. Beaming light from a subwavelength aperture. Science **297**, 820-822 (2002).

[24] Barnes, W. L., Murray, W. A., Dintinger, J., Devaux, E. & Ebbesen, T. W. Surface Plasmon Polaritons and Their Role in the Enhanced Transmission of Light Through Periodic Arrays of Subwavelength Holes in a Metal Film. Phys. Rev. Lett. **92**, 107401-107401 (2004).

[25] Agrawal, A., Matsui, T., Vardeny, Z. V. & Nahata, A. Terahertz transmission properties of quasiperiodic and aperiodic aperture arrays. J. Opt. Soc. Am. B **24**, 2545-2555 (2007).

[26] Seo, M. A. et al. Terahertz field enhancement by metallic nano slit operating beyond the skin-depth limit. Nature Photon. **3**, 152-156 (2009).

[27] Park, H. R. et al. Terahertz nanoresonators : Giant field enhancement and ultrabroadband performance. Appl. Phys. Letters **96**, 121106 (2010).

[28] Tanaka, H. et al. Enhancement of THz field in a gap of dipole antenna. 35[th] International Conference on Infrared Millimeter and Terahertz Waves (IRMMW-THz), 1-1 (2010).

[29] Novotny, L. & van Hulst N. Antennas for light. Nature Photon. **5**, 83-90 (2011).

[30] Feurer, T., et al. Terahertz polaritonics. Annu. Rev. Mater. Res. **37**, 317–350 (2007).

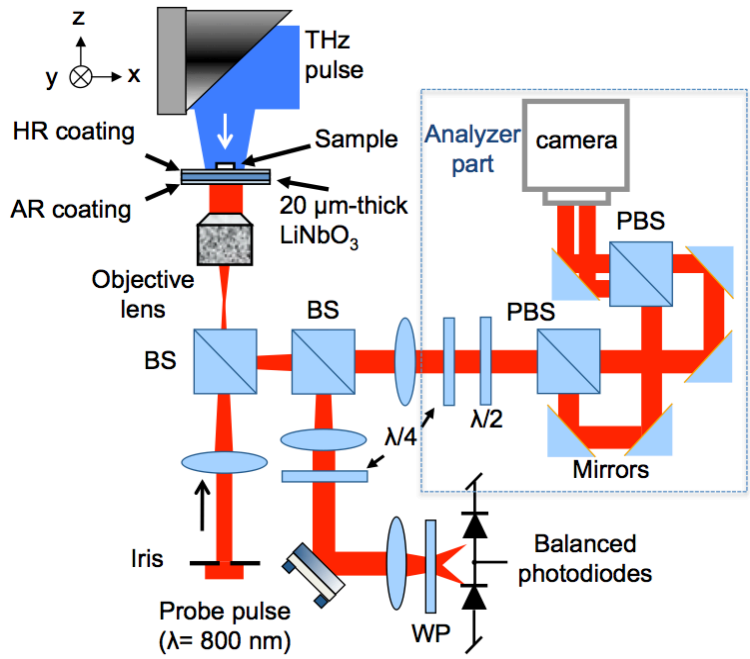

**Figure 1 Schematic of experimental setup**: BS: non-polarized beam splitter, PBS: polarized beam splitter, WP: Wollaston prism.

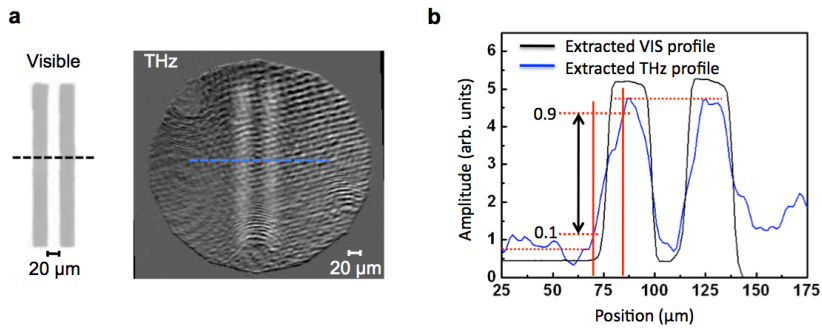

**Figure 2 Evaluation of spatial resolution**: **a**, Visible and THz images of a metallic mask. **b**, Extracted visible and THz profiles of metallic mask shown in (a). Spatial resolution of ~14 μm was found for time-resolved THz image. Extracted profile of visible image confirms that optical elements do not restrict spatial resolution for THz imaging.

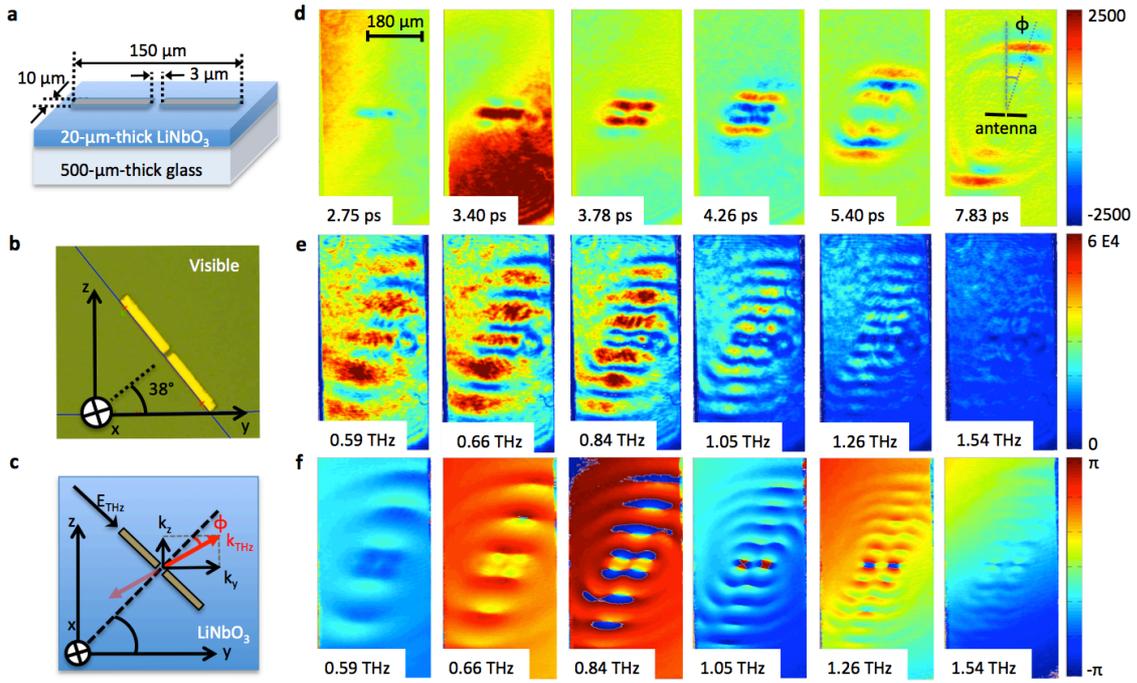

**Figure 3 Real-time THz near-field imaging**: **a**, Schematic and **b**, visible representation of dipole antenna **c**, Direction of the antenna relative to the LN crystal axis. **d**, Temporal snapshots of THz electric field. **e**, Amplitude and **f**, Phase components of Fourier transforms obtained from time-dependent field distribution for each pixel.

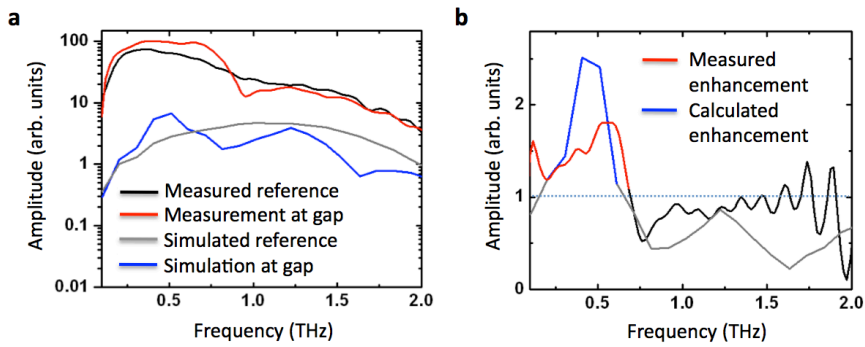

**Figure 4 Comparison between simulated and experimental enhancement fields at gap position of dipole antenna**: **a**, Fourier transformed of simulated and measured THz pulses at gap position with and without antenna. **b**, Ratios of spectrums presented in (a).

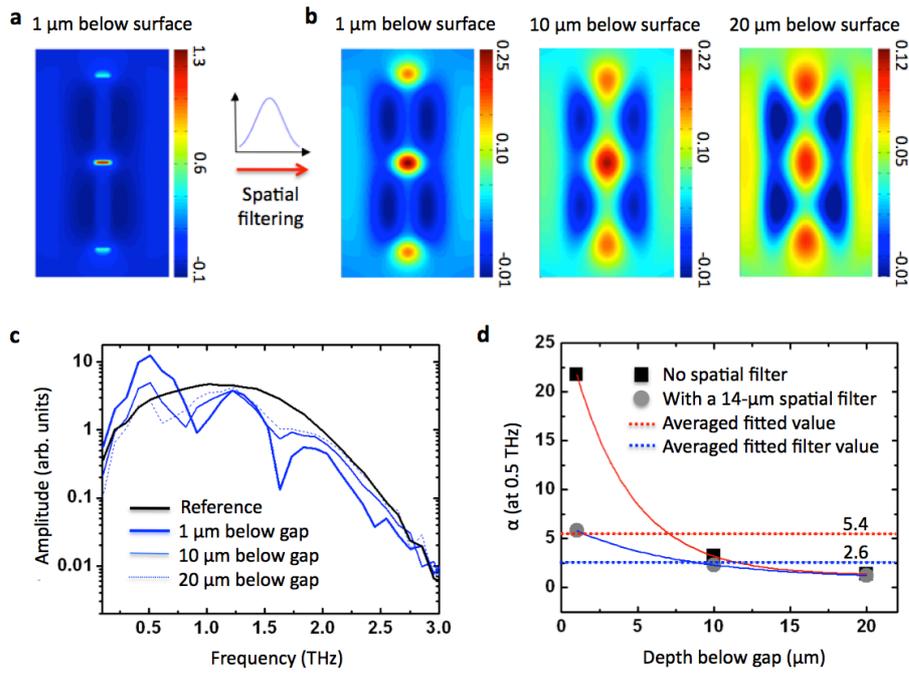

**Figure 5 Simulated electric field mapping of dipole antenna**: **a,** Mapped field 1 µm below LN crystal surface with a mesh resolution of 1 x 1 µm. (Antenna is oriented vertically) **b**, Mapped field as a function of depths below crystal surface including a Gaussian low pass filter that lowered the spatial resolution to 14 µm. **c**, Fourier transformed of simulated reference THz pulse with Fourier transformed of THz field at gap position as a function of z direction. **d**, Peak electric field at 0.5 THz normalized with THz pump reference as a function of z direction. Red and blue dotted lines are average enhancement factors for 1-µm and for 14-µm spatial resolutions.